\documentclass[12pt,preprint]{aastex}

\usepackage{graphicx}

\begin{document}


\title{Contemporaneous {\it XMM-Newton} investigation of a giant X-ray flare and 
quiescent state from a cool M-class dwarf in the local cavity}

\author{ A. Gupta and M. Galeazzi\altaffilmark{1}}
\affil{Physics Department, University of Miami, Coral Gables, FL 33124, USA}
\altaffiltext{1}{Corresponding author, galeazzi@physics.miami.edu}

\author{B. Williams}
\affil{Astronomy Dept., University of Washington, Box 351580, Seattle WA 98195, USA}

\begin{abstract}

We report the serendipitous detection of a giant X-ray flare from the source 
2XMM J043527.2-144301 during an {\it XMM-Newton} observation of the high latitude 
molecular cloud MBM20. The source has not been previously
studied at any wavelength. The X-ray flux increases by a factor of more than 52 from 
quiescent state to peak of flare. A 2MASS counterpart has been identified 
(2MASS~J04352724-1443017), and near-infrared colors reveal a spectral 
type of M8-M8.5 and a distance of $(67\pm 13)$~pc, placing the source in front
of  MBM20. Spectral analysis and source luminosity are also consistent 
with this conclusion. The measured distance makes this object
the most distant source (by about a factor of 4) at this spectral type 
detected in X-rays.
The X-ray flare was characterized by peak X-ray luminosity of 
$\sim 8.2\times 10^{28}$~erg~s$^{-1}$ and integrated X-ray energy of 
$\sim 2.3 \times 10^{32}$~erg. 
The flare emission has been characterized with a 2-temperature model with 
temperatures of $\sim$10 and 46~MK (0.82 and 3.97~keV), and is 
dominated by the higher temperature component. 

\end{abstract}

\section{Introduction}

In the solar neighborhood a significant amount of main sequence stars are 
cool stars of spectral class M. Stars with spectral type M7 and later, 
sometimes called ``ultracool M dwarfs'', are optically very faint, making 
them difficult to identify in optical, near-infrared and high proper motion 
sky surveys. They may, however, be characterized by strong X-ray flares.

X-ray flares can provide clues on the size of the X-ray emitting structures 
and on the underlying physical processes that produce very
energetic events. The soft X-ray light curves of stellar 
flares are typically characterized by a fast rise phase,
followed by a slower decay, and the analysis of 
the decay phase can be used to diagnose the flaring loops length 
\citep{Reale2002,Reale2007}. 

A sudden increase of temperature and emission measure of solar and stellar 
coronal flares is caused by strong heat pulses of the plasma confined in 
single or groups of magnetic loops.  The flare in ultracool dwarfs suggests 
a presence of magnetic activity in the outer atmospheric layers of these stars.
In massive solar type stars, magnetic activity is generated at the interface 
layer between the radiative core and outer convection zone. However, around 
spectral type M5/M6 or later, the stellar interior is fully convective and 
short-term enhancements of the emission in various wavebands are
attributed to flares due to magnetic reconnection.
\citep{Hambaryan2004, Stelzer2006, Robrade2009}. 

Flares in active late K and M dwarfs commonly show short lived events 
with typical duration of minutes to tens of minutes and integrated X-ray 
energies of $\sim10^{30}$ to $10^{33}$~erg \citep{Pallavicini1995}.
However, occasionally intense long duration events lasting from
a few to several hours are also observed from M dwarfs stars
\citep{Pallaviciin1990,Singh1999,Wargelin2008, Laycock2009}.
Many M dwarfs are also known to show X-ray flares with flux increases
up to a factor 100 to 200 (Robrade \& Schmitt 2009, and references therein).
Nearly all late type stars emit X-rays believed to originate
in $10^{6.0}$ to $10^{7.5}$~K plasma confined in closed magnetic
structures \citep{Singh1999,Giampapa1996}.

The X-ray spectrum of these objects is commonly modeled as two 
thermal components corresponding to two distinct types of loop atmosphere.
The low temperature component, or quiet region, has temperature around 
(2-4)$\times 10^6$~K, while the high temperature component, 
or active region, has typical temperature around $10^7$~K 
\citep{Kashyap1992, Giampapa1996, Fleming2003}. It has been observed that, 
in active M dwarfs, the hot component contributes a relatively larger fraction of the
total X-ray emission and is also responsible for the variability in the X-ray 
light curve \citep{Fleming2003}. The low temperature component dominates the
 X-ray emission in inactive dwarfs 
with little or no variability in the light curves.

As part of a campaign to characterize the properties of the diffuse
x-ray background \citep{Galeazzi2007,Galeazzi2009,Gupta2009} we performed a 
100~ks {\it XMM-Newton} observation of the nearby star forming cloud MBM20
to a flux limit of $5.5\times10^{-16}$~ergs~cm$^{-2}$~s$^{-1}$
in the energy band (0.5-2)~keV. Several bright point sources
were detected during the observation, most of them background active 
galactic nuclei, but we also observed some foreground (Galactic) stars. 

One of the sources identified (2XMM~J043527.2-144301) was below
the threshold for source detection during the first half 
of the observation, becoming one of the brightest sources in
the field during the second half. A detailed analysis of the data
showed that this was due to two consecutive X-ray flares from the source 
happening toward the end of the observation. With the source location 
obtained from the second half of the observation, we also verified 
that the source was actually visible, although below our
threshold for detection, during the first half of the observation,
providing the rare opportunity of observing, in the same data set, 
both quiescent state and the two flares. 
By comparing the sources location with available catalogs, we were also 
able to identify the source as a cool M-class dwarf in the solar neighborhood,
with a faint 2MASS counterpart (2MASS~J04352724-1443017).

In this paper we detail our analysis of the flares and quiescent state
from the X-ray source, and the corresponding 2MASS counterpart.   
In section 2 we describe the observation and data reduction, while 
in section 3 we focus on the data analysis. In particular, in section 3.1 
we discuss the timing analysis of the X-ray flares, section 3.2 is dedicated 
to the analysis of the optical and NIR counterpart, and section 3.3 to the 
spectral analysis of the X-ray source. A summary and conclusions are 
reported in section 4. 

\section{Observation and data reduction}

MBM20 was observed by {\it XMM-Newton} in August 2004 for approximately
$100$~ks (ObsID 0203900201). MBM20 is one of the closest star forming 
clouds \citep{Russeil2003} and is probably located within or at the edge 
of the Local Cavity. Its mass is $\sim 84$~M$_\odot$, and it has coordinates 
$l=211^\circ23'53''.2$, $b=-36^\circ32'41''.8$, southwest of the Orion 
star-forming complex.  The currently accepted distance to \object{MBM20} is
$112\pm15$~pc~$<d<161\pm21$~pc \citep{Hearty00}.
The evaluated neutral hydrogen density in the direction of the denser part 
of MBM20 is $1.59\times10^{21}$~cm$^{-2}$.
The primary aim of this observation was to study the properties of the diffuse 
X-ray background \citep{Galeazzi2007,Galeazzi2009}. During the analysis 
of the data, we observed a large flare from source 2XMM~J043527.2-144301 during the
last $30$~ks of the $100$~ks exposure,
while in the first $40$~ks of the exposure the source was in its
quiescent state.

Data analysis was carried out with the standard {\it XMM} software, the
Science Analysis System (SAS) version 10.0.0. Although {\it XMM-Newton} 
carries three imaging detectors, the European Photon Imaging Cameras
(EPIC) \emph{pn}, \emph{MOS1} and \emph{MOS2}, the source falls in the CCD gap of the 
MOS detectors, thus we considered only X-ray data taken with the 
\emph{pn} detector.

The XMM-SAS task \emph{xmmselect} was used to extract the light-curves 
and spectra of the source and background. 
We selected a circular region of radius 25\arcsec~around the source 
position to extract the light-curve and spectra of the source (Fig.~1). 
To extract the background light-curve and spectra we used two different 
approches (regions). First we defined the background region 
as a 75\arcsec~radius circular ring around the source region. The region 
contains an additional point source that was removed. 
As an alternative, we used a close-by region, on the same CCD
that contains the source, with the same dimensions of the source region. 
We did not find any significant difference in the background from the two 
methods. We extracted the source and background light-curves 
and spectra, restricting to events spread at most in two contiguous 
pixels (i.e., $pattern=0-4$) and to the energy range from 0.2 to 7.5~keV. 
Further we used SAS task  \emph{epiclccorr} to subtract background from the 
source light-curve. We also generated exposure maps to account for 
spatial quantum efficiency, mirror vignetting, and field of view by
running SAS task \emph{eexpmap}.

\section{Analysis}

\subsection{Timing Analysis}

Figure~2 shows the background-subtracted X-ray light curve with
1~ks binning for source 2XMM~J043527.2-144301. 
The source light curve clearly shows two flares 
that occurred after an initial quiescent period of at 
least 65~ks (including a time interval in the middle of 
the observation affected by proton flare). The source first 
flare was very strong; in the energy range (0.2-7.5)~keV 
the peak count rate reached in the light curve was 
(0.11-0.12)~counts~s$^{-1}$, depending 
on the choice of binning. However 
during the quiescent period the source was 
very faint, with an average count rate of 
0.0021~counts~s$^{-1}$, corresponding to an increase of more 
than 52 in the count rate from quiescent to peak. The second 
flare in the same energy band had peak count rate of 
0.055~counts~s$^{-1}$.  Both flares 
show fast-rise, exponential-decay profile with a rise time 
of $\sim$3~ks and similar e-folding decay times 
of $(4.3\pm0.7)$~ks and $(4.5\pm1.0)$~ks for the first and second 
flare respectively.

\subsection{Optical/Near-IR Counterpart}

We compared the {\it XMM-Newton} position of source 
2XMM~J043527.2-144301 with error radius of 4\arcsec~ with available optical/IR catalogs.
We found a faint 2MASS counterpart (2MASS~J04352724-1443017)
only 0.48\arcsec~ from the {\it XMM-Newton} position, with $K_{s}=14.35\pm0.08$
and $(J-K_{s})=1.16\pm0.10$ (Fig 3, Table 1). There is only one 
source in a $4\arcsec$ radius in the near-infrared band of DSS image,
which is not visible at other wavelengths (Fig.~3).

We also found a counterpart of our source in the USNO-B1.0 
catalog and Guide Star Catalog GSC-2.3 with offset of 0.9\arcsec. 
The USNO-B1.0 counterpart has R-band magnitude of 19.37 and near-IR
band magnitude of 17.9. Whereas GSC counterpart is identified only 
in near-IR ($0.8\micron$  band), no detection in F (red), J (blue) and
V (green) photographic band. GSC-2.3 catalog is based on
ground-based photographic plate material. The majority of
stars visible in these cameras plates are very red late-type
stars, in agreement with the 2MASS characterization (see next paragraph).

To assign a spectral class to source 2XMM~J043527.2-144301,
we compared the 2MASS colors of this source with the colors
of other M dwarfs\footnote{$http://www.pas.rochester.edu/\sim emamajek/memo\_colors.html$} 
and found that the $(J-H)$, $(H-K_{s})$ and $(J-K_{s})$ colors are 
best matched the spectral class M8.5V. To confirm the spectral class, 
in Fig.~4 we compare source 2XMM J043527.2-144301 with data from
Gizis et al. (2000, their Fig.~4) on a color-color diagram. In particular,
the figure shows typical $(J-H)$ vs. $(H-K_{s}$)
value for M7-M7.5 (horizontal stripes region) and M8 or
later (vertical stripes region) dwarfs. Most M7-M7.5 dwarfs
are around $(H-K_{s}, J-H)=(0.45, 0.62)$ whereas 
M8-M8.5 dwarfs are around $(H-K_{s}, J-H)=(0.42,0.74)$
with similar $J-K_{s}$ colors (although, as the vertical striped region
indicates, a few M8 or later stars may have singificantly higher numbers).
The black star in the figure shows the colors 
$(H-K_{s}, J-H)=(0.42\pm0.10,0.74\pm0.07)$ of the 2MASS counterpart 
of the source 2XMM~J043527.2-144301, indicating a spectral class 
of M8-M8.5. We note that to properly confirm the spectral type of the source 
dedicated spectroscopic observations are necessary.

To determine the distance of the source, we estimated
the absolute magnitude from the 2MASS $(J-K_{s})$ color using the relation 
from Gizis et al. (2000), $M_{k}=7.593+2.25\times(J-K_{s})$,
with a scatter of $\sigma=0.36$ magnitudes. The relation is valid
 for only M7 and later dwarfs over the color range.
The source 2XMM~J043527.2-144301 value of $(J-K_{s})=(1.16\pm0.10)$
gives $M_{k}=10.21\pm0.22$, and using the standard distance
modulus equation $(m_{k}-M_{k}=5\log \frac{d_{k}}{10})$,
we get  $d_{k}=(67.\pm13)$~pc. We note that this distance
is significantly shorter than the measured distance of MBM20 
$(112\pm15$~pc~$<d<161\pm21$~pc), indicating that source 
is not part of MBM20, but in front of it.
As discussed in the next sessions, this is consistent with the 
spectral analysis of the flares and their inferred luminosity.
In our subsequent analysis we use the distance derived from the
2MASS  $(J-K_{s})$, 
although we will also briefly discuss the possibility that, instead,
source 2XMM~J043527.2-144301 is part of MBM20.

Using the relation for Bolometric Correction $(B.C.)_k$  from 
Legget (2001), we also obtained a bolometric magnitude of 
$M_{bol} = (13.32\pm 0.43)$, which leads to a bolometric 
luminosity of $L_{bol}=(1.4\pm0.6)\times 10^{30}$~erg~s$^{-1}$. 
We then used the near-IR colors to estimate the radius, effective 
temperature and mass of the source 2XMM J043527.2-144301.
Assuming a typical ultracool dwarfs age of $\tau<$2 to 3~Gyrs, we used the 
relation from Dahn et al.(2002), $R/R_{\odot}=0.088+0.00070(16.2-M_{bol})^{2.9}$,
which predics radii for objects with age between 1~Gyr and 5~Gyr 
over the range 12-16.5 in $M_{bol}$, to calculated the radius of the 
source:
\begin{equation}
\frac{R}{R_{\odot}}=0.103\pm0.006.
\end{equation}

With the calculated values of luminosity and radius, we also determined the 
effective temperature of the source as $T_{eff}=(2479\pm275)$~K, 
which is typical for an ultracool M dwarf. 
Using the relationships for Mass vs. M$_k$ 
and Mass vs. spectral-type from Chabrier et al. 
(2000 - their Figs. 9~\&~10 respectively), we found that the mass of 
2XMM~J043527.2-144301 is in the range 
$(0.079 M_{\odot} - 0.083 M_{\odot})$, suggesting that the object 
is near the stellar/substellar transition.

\subsection{Spectral Analysis}

To characterize the process responsible for the X-ray emission
from source 2XMM J043527.2-144301, we analyzed its \emph{pn}
spectra during the quiescent and flare states. 
Source and background 
spectra in the (0.2-7.5)~keV band were produced with SAS task
\textrm{$xmmselect$}, using the same regions used to extract the
light-curve. The SAS tasks \textrm{$rmfgen$} and \textrm{$arfgen$}
were used to produce redistribution matrices files (RMF) and 
ancillary response files (ARF).

Spectral fitting of the background subtracted spectrum was
performed with XSPEC~v~12.4. We applied standard stellar emission models 
such as absorbed 1-temperature (1T) and 2-temperature (2T) thermal
plasma models.
For plasma thermal emission, the APEC \citep{Smith2001} model 
was used. We also repeated the fits with other spectral models for
thermal equilibrium plasma such as Raymond-Smith (RAYMOND) and
Mewe-Kaastra-Liedahl (MeKaL), but we did not find any 
significant change in the fit parameters or reduced $\chi^{2}$.

During the quiescent state the source was very faint,
with net counts (after removing background) of $(40\pm14)$
in 18990~s (exposure corrected time) in the energy range 
$(0.2-7.5)$~keV, for a quiescent count rate of 
$(2.1\pm0.7)\times10^{-3}$~counts~s$^{-1}$.
For model fitting, the source spectrum (during the quiescent state)
was grouped to have 10 counts per channel (before background subtraction). 
The spectrum is well fitted (Fig. 5) by an absorbed cool
thermal plasma model, however, the value of the temperature
varies significantly depending on the binning and interval used 
for the fit. 
Combining all results, the best value 
for the temperature is $kT=(1.7\pm0.8)$~keV, while the
absorption is poorly constrained. Due to the poor constraints on
the fit with the 1T model, no significant improvement was obtained
using a 2T plasma model. We note that the spectrum is also affected 
by an excess of counts at low energy, probably due to residual 
background. 

During the flare, 2XMM~J043527.2-144301 was one of the
brightest sources in the field. A weaker flare was also
observed during the decay of the first giant flare. We analyzed the source
spectrum during the first (strong) and second (weak) flares separately
and didn't find any significant difference between the two. 
We therefore only report here the specral analysis of the sum of the two.

The first flare occurred during the observing time
70.0 to 80.0~ks (exposure corrected
time of 6.14~ks). The source had net counts of $(267\pm16)$ and
$(276\pm17)$ in the energy bands $(0.2-2.0)$~KeV
and $(0.2-7.5)$~keV respectively, corresponding to average
count rates of $(4.3\pm0.3)\times10 ^{-2}$~counts~s$^{-1}$ and
$(4.5\pm0.3)\times10^{-2}$~counts~s$^{-1}$. 
The second flare had net counts of $(90\pm9)$ for an exposure 
corrected time of 3.35~ks (observing time 80.0 to 86.0~ks) 
in the energy range of 0.2-2.0~keV, corresponding 
to a count rate of $(2.7\pm0.3)\times10^{-2}$~counts~s$^{-1}$.

Since most of the emission from the source is in the energy
range of 0.2-2.0~keV, as expected from late type dwarf,
we limited our spectral fitting to the 0.2-2.0~keV energy range.
We grouped the spectrum to have 15 counts per channel. The 1T thermal
plasma models gave unacceptably high values for reduced chi square
($\chi^{2}=63$, $n=16$), however, the spectral model with 2T thermal 
plasma resulted in better fits.
The fit results are summarized in Table 2 and fits
are shown in Fig.~5. In the fits, the absorption is still poorly 
constrained, but is consistent with very low neutral hydrogen column density (NH).

To set a limit on NH, we repeated the fits with both free and fixed absorption. 
The fits with free absorption
were statistically compatible to those with NH fixed at zero.
We also repeated the fits by increasing the value of the fixed 
NH. By looking at the quality of the fit (value of $\chi ^2$), we were able to
set an upper limit on the neutral hydrogen column density of 
NH$<2.8\times10^{20}$~cm$^{-2}$. Note that the effect of the varition in 
NH on the other fit parameters is reflected in the results reported in Table~2. 
Using IRAS100~\micron~data 
\citep{Galeazzi2007} we estimated the neutral hydrogen
column density of MBM20 in the same direction to be 
$1.7\times10^{21}$~cm$^{-2}$, supporting the notion that the source is, 
indeed, in front of MBM20.

We also extracted the spectrum of the peak emission of the 
first flare. We defined the peak period as the interval in which 
the count rate was above 50\% of the maximum value (i.e. greater
than 0.057~counts~s$^{-1}$). For the peak emission the 
exposure corrected time is 2.79~ks (from 71.2 to 75.4~ks).
The 2T thermal plasma model results in good fit (Table~2) with
the temperature of the cooler component similar to that obtained from the
sum of the two flares; however the temperature of the hotter
component increases from $\sim$29~MK to $\sim$46~MK.

\subsection{Energy budget}

During the quiescent period the flux of the source in the
energy band (0.2-2.0)~keV was
$(9.1\pm2.5)\times10^{-15}$~erg~s$^{-1}$~cm$^{-2}$,
where uncertainties are 90\% confidence range.
Assuming the distance $d_{k}=(67\pm13)$~pc,
inferred from 2MASS colors, this leads to a luminosity
of $(4.9\pm2.4)\times10^{27}$~erg~s$^{-1}$. 

The average flux of the source during the first flare in the energy
band $(0.2-2.0)$~keV was 
$(1.07\pm0.14)\times10^{-13}$~erg~s$^{-1}$~cm$^{-2}$, and the corresponding 
luminosity $(5.8\pm2.4)\times10^{28}$~erg~s$^{-1}$. 
Integrating the X-ray luminosity during the time of the
first flare, including the fraction overlapping with the second flare,
the total emitted X-ray energy was $(6.7\pm2.8)\times10^{32}$~erg.

For the peak emission of the first flare,
the flux in the energy band $(0.2-2.0)$~keV was
$(1.5\pm0.4)\times10^{-13}$~erg~s$^{-1}$~cm$^{-2}$ and the
luminosity $(8.2\pm3.7)\times10^{28}$~erg~s$^{-1}$, while
the total X-ray energy released during peak emission
was $(3.5\pm1.5)\times10^{32}$~erg.
Using the 2T thermal plasma model parameters, the count
rate of the flare at the peak ($\sim0.12$~counts~s$^{-1}$)
can also be converted to a flux of 
$(2.8\pm0.4)\times10^{-13}$~erg~s$^{-1}$~cm$^{-2}$
 and corresponding luminosity of $(1.5\pm0.6)\times10^{29}$~erg~s$^{-1}$.

The average flux of the second flare in the energy band 
$(0.2-2.0)$~keV was $(6.8\pm2.1)\times10^{-14}$~erg~s$^{-1}$~cm$^{-2}$ 
and the corresponding luminosity
$(3.7\pm1.8)\times10^{28}$~erg~s$^{-1}$. Removing the contribution
of the first flare, the net energy released from the second flare 
was $(1.3\pm1.1)\times10^{32}$~erg. 
Notice that the total X-ray energy released during the first flare was 
about ~6 times the total X-ray energy released during the second one. 
The total X-ray energy emitted by the source through both flares was 
$(7.7\pm3.3)\times10^{32}$~erg.

We also compared the X-ray and bolometric luminosity of source 
2XMM~J043527.2-144301 to find its activity level. We found the log of
the ratio to be $\log{L_x /L_{bol}}=(-2.5\pm0.7)$ in the quiescent state 
and $\log{L_x /L_{bol}}=(-1.0\pm0.6)$ at the flare peak, indicating a 
high activity level of the flare. These values are comparable to
what has been observed by Hambaryan (2003) for M9 dwarf 
IRXS~J115928.5-524717.

We compared our results with previous investigations of other
ultracool M dwarfs. Table~3 summarizes the luminosities and
coronal temperatures of all the observations available in X-rays.
The luminosity of ultracool M dwarfs in low activity
period varies from $2.5\times10^{25}$~erg~s$^{-1}$ to
$2.0\times10^{27}$~erg~s$^{-1}$. Our value for luminosity 
$(4.9\pm2.4)\times10^{27}$~erg~s$^{-1}$ is close to the upper range 
of luminosities observed so far in the quiescent state for ultracool 
M dwarfs. The same is true for the flare mean and peak emission 
luminosities.

The luminosities and coronal temperatures of our source in
the quiescent and flaring periods are of the same order as observed for
M8 dwarf LP 412-31. Stelzer et al. (2006) observed a 
giant flare from LP 412-31 lasting for $\sim$2~ks with peak luminosity of 
$4.7\times10^{29}$~erg~s$^{-1}$. They reported an increase in the
count-rate by a factor 200-300 from quiescent to peak of flare. The 
emitted flare energy in soft X-rays was $3.0\times10^{32}$~erg,
similar to the value we observed for 2XMM~J043527.2-144301. 

We note that, if instead of the measured distance the source is 
associated with MBM20, based on the cloud distance all the values 
reported would be at least three times bigger. This would be 
inconsistent with data from previous investigations, 
strengthening the conclusion that, indeed, the source
is not part of MBM20, but in front of it.

\section{Summary and conclusions}

\begin{enumerate}
\item We detected a giant flare as well as the quiescent X-ray emission 
from source 2XMM~J043527.2-144301. This is the first analysis report of this source in 
X-rays or any other wavelength.

\item During the decay of the giant flare, a second, weaker flare was also observed.
Both flares show a fast-rise-exponential-decay 
profile with a rise time of $\sim$3~ks, and similar 
e-folding decay times of $(4.3\pm0.7)$~ks and $(4.5\pm1.0)$~ks
respectively.  

\item In the quiescence state the source is detected with
an X-ray luminosity of $\sim 4.9\times10^{27}$~erg~s$^{-1}$ and the
emission can be characterized by a single temperature
at $\sim$20~MK. The X-ray flare shows an intensification of 
more than 52 times from quiescent state to flare peak, with a peak 
X-ray luminosity of $\sim 1.5\times10^{29}$~erg~s$^{-1}$. 
The flare emission can be characterized by a two temperature emission
at $\sim$10 and 46~MK and is dominated by the higher temperature component. 

\item We have identified an optical/near-IR counterpart to source 
2XMM~J043527.2-144301. Its characterization through 
optical/Near-IR colors suggests the source as late-type star, 
most probably an ultracool dwarf of spectral type M8-M8.5. 
Although beyond the scope of this paper, further spectroscopic observations 
are encouraged to confirm the spectral type of the source.

\item The X-ray spectrum of source 2XMM~J043527.2-144301
 shows similarities with the spectra of other late type dwarfs,
and the temperatures of 1T and 2T thermal plasma models
 and X-ray luminosities of quiescence and flare 
states of the source are comparable
 to other late type ultracool dwarf spectral models
 \citep{Hambaryan2004,Robrade2009,Robrade2010}.

\item Only a very small number of X-ray observations of ultracool 
dwarfs are mentioned in the literature, and source 
2XMM~J043527.2-144301 is one of the few 
ultracool dwarfs exhibiting a giant X-ray flare that 
is also observed during the quiescent state.
The estimated distance of $(67\pm13)$~pc, also makes this object 
the most distant source (by about a factor of 4) at this spectral type 
detected in X-rays.

\end{enumerate}

\acknowledgments
 This investigation was supported in part by NASA grant \# NNG05GA84G.
 We have made use of the data products from the Two Micron All Sky Survey,
 which is a joint project of the University of Massachusetts and the
 Infrared Processing and Analysis Center/California Institute of Technology, 
funded by the National Aeronautics and Space Administration and 
the National Science Foundation. We have also
 used the USNOFS Image and Catalog Archive operated by 
the United States Naval Observatory,
 Flagstaff Station and the Digitized Sky Surveys 
produced at the Space Telescope Science Institute
 under U.S. Government grant NAG W-2166.

\clearpage

\begin{figure}
\epsscale{1}
\begin{center}
\includegraphics[width=15cm]{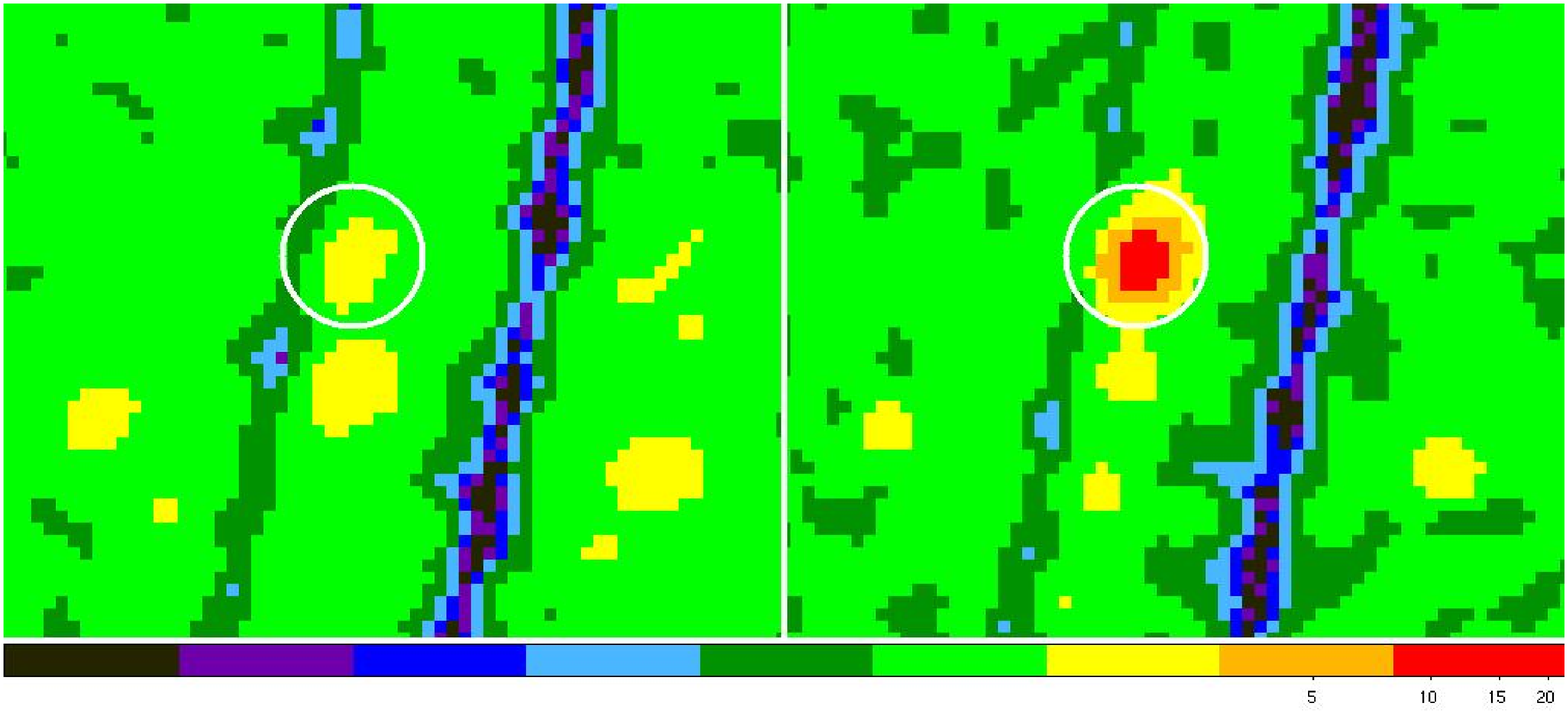}
\end{center}
\caption{{\it XMM-Newton pn} image of the MBM20 field containing 
the source 2XMM~J043527.2-144301, during quiescence 
({\it left}) and flare ({\it right}). A 25\arcsec~ circular 
region around the source position was used to extract light curves 
and spectra (white circles).}
\end{figure}

\clearpage

\begin{figure}
\epsscale{1}
\begin{center}
\includegraphics[width=15cm]{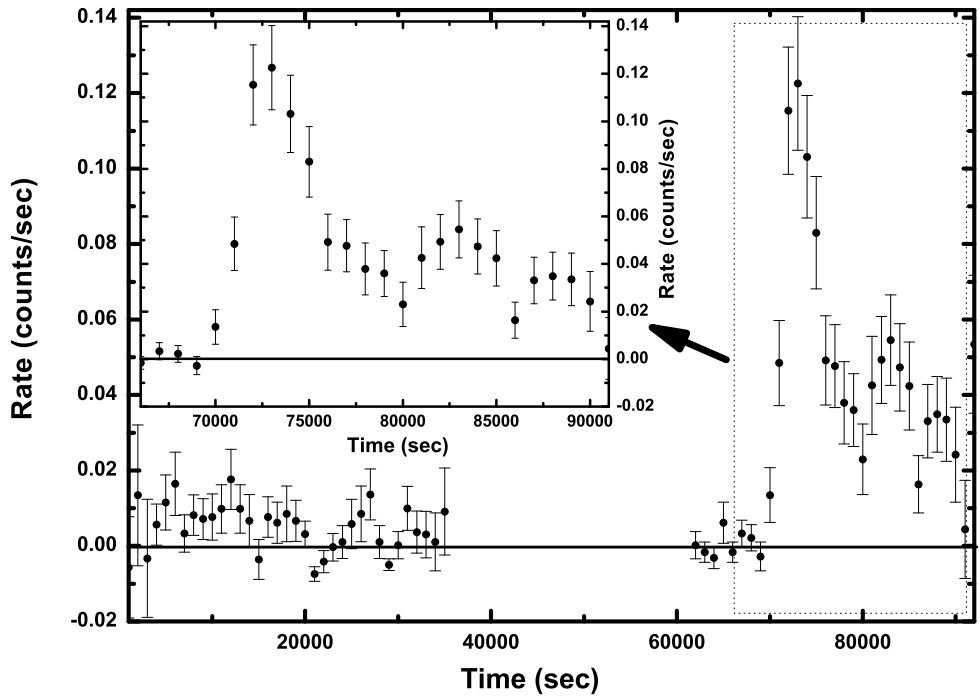}
\end{center}
\caption{{\it XMM-Newton pn} background-subtracted 0.2-7.5~keV light curve of 
source 2XMM~J043527.2-144301, binned at 1~ks resolution. 
The central region without datapoints was affected by strong proton flares, 
therefore excluded from our analysis.}
\end{figure}

\clearpage

\begin{figure}
\epsscale{.80}
\begin{center}
\includegraphics[width=15cm]{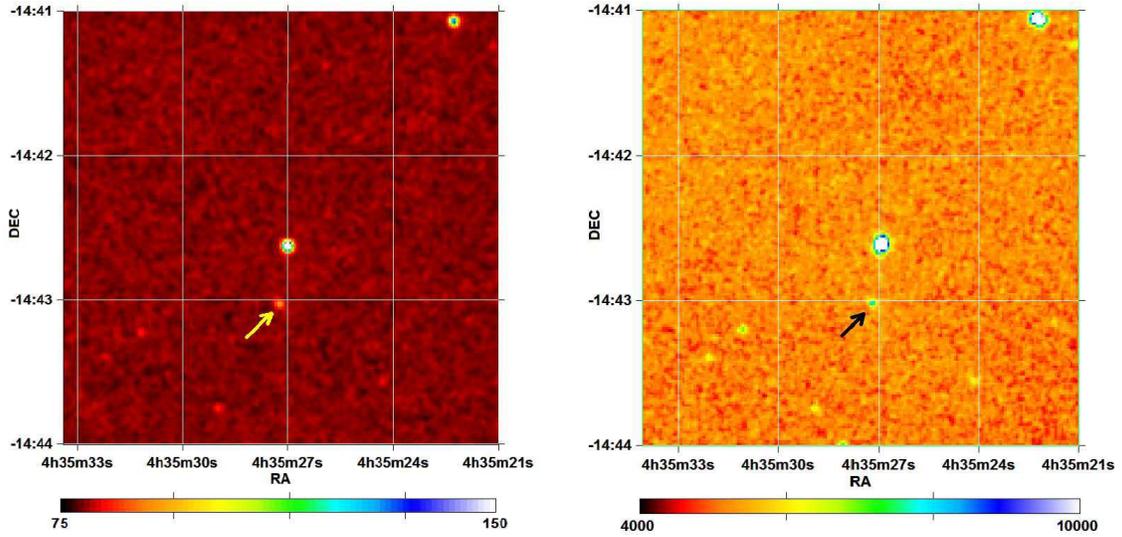}
\end{center}
\caption{Images in 2MASS $K_{s}$ band ({\it left}) and DSS image in
 near-infrared band ({\it right}) of the MBM20 field.
The arrows shows the location of the 2XMM~J043527.2-144301 counterpart.}
\end{figure}

\clearpage

\begin{figure}
\epsscale{1}
\begin{center}
\includegraphics[width=10cm]{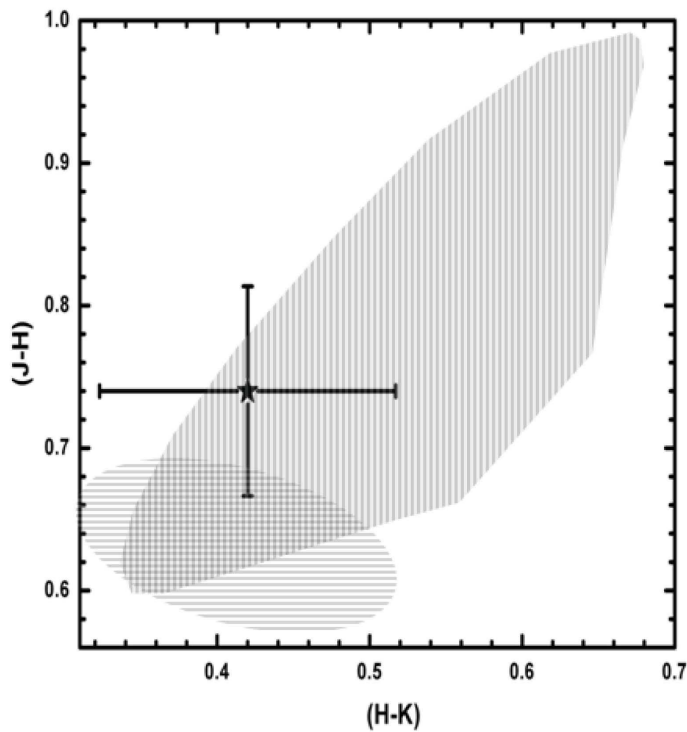}
\end{center}
\caption{2MASS near-infrared color-color diagram using data
 from Gizis et al. (2000). M8 and later dwarfs are represented
 by the vertical striped region, while M7-M7.5 dwarfs are
 represented by the horizontal striped one. 
Most M7-M7.5 dwarfs are around $(H-K_{s}, J-H)=(0.45, 0.62)$ 
whereas M8-M8.5 dwarfs are around $(H-K_{s}, J-H)=(0.42,0.7)$
with similar $J-K_{s}$ colors (although, as the vertical striped region
shows, a few M8 or later stars may have singificantly higher numbers).
The data-point represents the colors $(H-K_{s}, J-H)=(0.42,0.74)$ of
the 2MASS counterpart of 2XMM~J043527.2-144301, indicating
a spectral class of M8-M8.5.}
\end{figure}

\clearpage

\begin{figure}
\epsscale{1}
\begin{center}
\includegraphics[width=15cm]{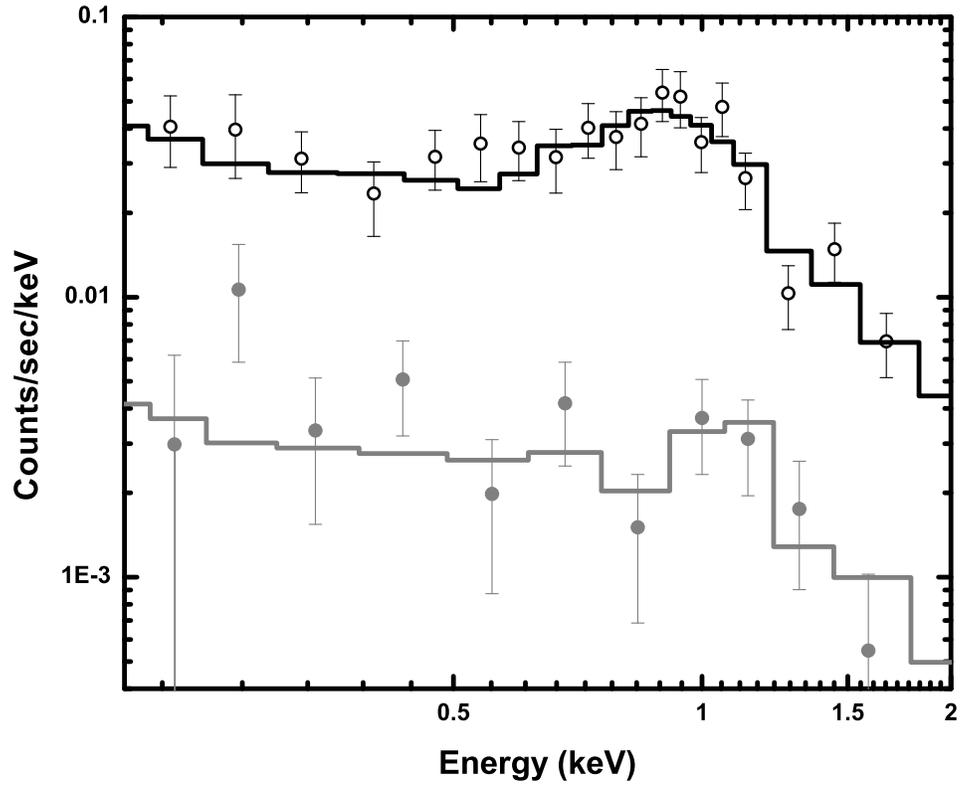}
\end{center}
\caption{{\it XMM-Newton pn} spectra of 2XMM~J043527.2-144301 during 
quiescent (filled grey circles) and flare time (empty black circles).
The solid lines show the fitted 1T (quiescent) and 2T (flare) thermal 
plasma models.}
\end{figure}

\clearpage

\begin{deluxetable}{ccc}
\tabletypesize{\scriptsize}
\tablecaption{Optical/NIR magnitudes of the 2XMM~J043527.2-144301 counterparts.}
\tablewidth{0pt}
\tablehead{
\colhead{Catalog} & \colhead{Band} & \colhead{Magnitude}
}
\startdata
 & $J_{mag}$ & $15.51\pm0.05$ \\
2MASS & $H_{mag}$ & $14.77\pm0.05$ \\
 & $K_{mag}$ & $14.35\pm0.08$ \\
GSC & $N_{mag}$ & $17.69$ \\
USNO-B1.0 & $I_{mag}$ & $17.9$ \\
\enddata
\end{deluxetable}

\clearpage

\begin{deluxetable}{lccc}
\tablecolumns{4}
\tablewidth{0pc}
\tabletypesize{\scriptsize}
\tablecaption{Spectral Fitting Results}
\tablehead{
\colhead{Parameters}  & \colhead{Quiescent}  &  \colhead{Flares} &  \colhead{Flare 1(peak)} \\
\colhead{} & \colhead{1T Model}  & \colhead{2T Model} & \colhead{2T Model}
}
\startdata
$kT_1$ (keV) & $1.7\pm0.8$ & $0.77\pm0.07$ & $0.82\pm0.07$ \\
$EM_1$ $(10^{50}$~cm$^{-3})$ & $4.0\pm3.3$ & $7.8\pm4.5$ & $15.9\pm9.7$\\
$kT_2$ (keV) & \nodata & $2.4\pm0.8$ & $4.0\pm2.0$ \\
$EM_2$ $(10^{50}$~cm$^{-3})$ & \nodata & $3.3\pm1.9$ & $48.\pm28.$\\
$\chi^2/d.o.f.$ & 7.0/8 & 11.3/14 & 5.2/8 \\
\cutinhead{X-ray Luminosities ($L_{x}$)}
$L_x$ $(10^{28}$~erg~s$^{-1})$ & $0.49\pm0.24$ & $5.4\pm2.2$ &$8.2\pm3.7$\\
\enddata
\end{deluxetable}

\clearpage

\begin{deluxetable}{lccccccc}
\rotate
\tablecolumns{8}
\tabletypesize{\scriptsize}
\tablewidth{0pc}
\tablecaption{Summary of parameters of ultracool M dwarfs}
\tablehead{
\colhead{}    &  \colhead{} & \multicolumn{2}{c}{$\log(L^{flare}_{x})$} & 
\colhead{$\log(L_{x}\tablenotemark{Q})$} & \colhead{$T\tablenotemark{Q}$} & \colhead{$T^{flare}$} & \colhead{ref} \\
\cline{3-4} \\
\colhead{Stra name} & \colhead{Spectral type} & \colhead{$\log(L^{mean}_{x})$} & 
\colhead{$\log(L^{peak}_{x})$} & 
\colhead{} & \colhead{T1/T2} & \colhead{T1/T2} \\
\colhead{} & \colhead{} & \colhead{(erg/s)} & \colhead{(erg/s)} & 
\colhead{(erg/s)} & \colhead{(keV)} & \colhead{(keV)}\\
}
\startdata
VB10 & M8 & 26.9 & $>27.4$ & 25.4 & $\thickapprox~0.26$ &
 $\thickapprox~1.3$ & $\mathbf1$ \\
TVLM 513-46546 & M8.5 & 25.1 & \nodata & \nodata & 
\nodata & 0.95 & $\mathbf2$ \\
IRXS-J115928.5-524717 & M9 & 28.1 & 
29.1 & 26.0/27.3 & 0.18/0.30 & \nodata  & $\mathbf3$, $\mathbf4$\\ 
LP 412-31 & M8 & \nodata & 29.7 & 27.2 & 0.30/1.29 &
 1.08-3.8\tablenotemark{1} & $\mathbf5$  \\
LHS 2065 & M9 & \nodata & \nodata & 26.3 & \nodata &
 \nodata & $\mathbf6$ \\
SCR 1845-6357 & M8.5/T5.5 & 27.3 & 27.7 & 26.1 &
 0.13/0.34 & 0.35/1.96 & $\mathbf7$ \\
 & & & & & & 0.61/2.46\tablenotemark{2} \\
 {\bf 2XMM J043527.2-144301} & M8.0-M8.5 & 28.8 & 28.9 & 27.7 & 1.7 & 0.74/2.79 &
 {\bf this paper} \\
 & & & & & & 0.82/4.01\tablenotemark{2} \\
\enddata
\tablenotetext{Q}{Quiescent} 
\tablenotetext{1}{Range of mean temperature of the 2T thermal model}
\tablenotetext{2}{Flare peak emission}
\tablerefs{(1) Fleming 2003; (2) Berger 2008; (3)Hambaryan 2004; (4) Robrade 2009;
 (5) Stelzer 2006; (6) Robrade 2008; (7) Robrade 2010; }
\end{deluxetable}

\end{document}